\documentclass[final,10 pt,times]{elsarticle}

\usepackage{graphicx}
\usepackage{dcolumn}
\usepackage{bm}
\usepackage{hyperref}
\usepackage{widetext}
\usepackage[mathlines]{lineno}

\begin{document}

\newcommand{\pT}{$p_T$}
\newcommand{\upctopo}{UPC\_Topo~}
\newcommand{\upcmain}{UPC\_Main~}
\newcommand{\oneN}{$1\rm{n}1\rm{n}$~}
\newcommand{\xN}{$\rm{Xn}\rm{Xn}$~}
\newcommand{\myro}{$\rho^{0}$~}
\newcommand{\auau}{\mbox{Au+Au}}        
\newcommand{\pbpb}{\mbox{Pb+Pb}}        
\newcommand{\dau}{d+{\rm Au}}
\newcommand{\piplus}{$\pi^{+}$}
\newcommand{\piminus}{$\pi^{-}$}
\newcommand{\pionpair}{$\pi^{+} \pi^{-}$}
\newcommand{\kplus}{$K^{+}$}
\newcommand{\kminus}{$K^{-}$}
\newcommand{\kaon}{$K$}
\newcommand{\pbar}{$\bar{\rm p}$}
\newcommand{\degree}[1]{$#1^{\circ}$}
\newcommand{\dbeta}{$1/\beta-1/\beta_{C}$}
\newcommand{\xF}{$x_{F}$}
\newcommand {\pp}{\mbox{p+p}}
\newcommand{\pythia}{\textsc{PYTHIA}}
\newcommand{\sqrts}{$\sqrt{s}$}
\newcommand{\snn}{$\sqrt{s_{_{\rm NN}}}$}
\newcommand{\dndy}{dN/d$y$}


\title{Coherent diffractive photoproduction of $\rho^{0}$ mesons on gold nuclei at RHIC}

\author{
L.~Adamczyk$^{1}$,
J.~K.~Adkins$^{19}$,
G.~Agakishiev$^{17}$,
M.~M.~Aggarwal$^{31}$,
Z.~Ahammed$^{50}$,
N.~N.~Ajitanand$^{40}$,
I.~Alekseev$^{15,26}$,
D.~M.~Anderson$^{42}$,
R.~Aoyama$^{46}$,
A.~Aparin$^{17}$
% ...
}

\address{$^{1}$AGH University of Science and Technology, FPACS, Cracow 30-059, Poland}
% ...
\address{$^{15}$Alikhanov Institute for Theoretical and Experimental Physics, Moscow 117218, Russia}
% ...
\address{$^{17}$Joint Institute for Nuclear Research, Dubna, 141 980, Russia}
% ...
\address{$^{19}$University of Kentucky, Lexington, Kentucky, 40506-0055}
% ...
\address{$^{26}$National Research Nuclear University MEPhI, Moscow 115409, Russia}
% ...
\address{$^{31}$Panjab University, Chandigarh 160014, India}
% ...
\address{$^{40}$State University Of New York, Stony Brook, NY 11794}
% ...
\address{$^{42}$Texas A\&M University, College Station, Texas 77843}
% ...
\address{$^{46}$University of Tsukuba, Tsukuba, Ibaraki, Japan,}
% ...
\address{$^{50}$Variable Energy Cyclotron Centre, Kolkata 700064, India}
% ...

\date{\today}

\begin{abstract}

The STAR Collaboration reports on the photoproduction of $\pi^+\pi^-$ pairs in gold-gold collisions  at a center-of-mass energy of 200 GeV/nucleon-pair.  These pion pairs are produced when a nearly-real photon emitted by one ion scatters from the other ion.   

We fit the $\pi^+\pi^-$ invariant mass spectrum with a combination of  \myro and $\omega$ resonances and a direct $\pi^+\pi^-$  continuum.  This is the first observation of the $\omega$ in ultra-peripheral collisions, and the first measurement of  $\rho-\omega$ interference at energies where photoproduction is dominated by Pomeron exchange.  The $\omega$ amplitude is consistent with the measured $\gamma p\rightarrow \omega p$  cross section, a classical Glauber calculation and the $\omega\rightarrow\pi^+\pi^-$ branching ratio.  The $\omega$ phase angle is similar to that observed at much lower energies, showing that the $\rho-\omega$ phase difference does not depend significantly on photon energy.

The $\rho^0$ differential cross section $d\sigma/dt$ exhibits a clear diffraction pattern, compatible with scattering from a gold nucleus, with 2 minima visible.  The positions of the diffractive minima agree better with the predictions of a quantum Glauber calculation that does not include nuclear shadowing than with a calculation that does include shadowing. 

\end{abstract}

\maketitle

\section{Introduction}

Relativistic heavy ions are accompanied by high photon fluxes due to their large electric charge and the strongly Lorentz contracted electric fields.  In relativistic heavy ion collisions, these fields can produce photonuclear interactions. When the nuclei collide and interact hadronically, strong interactions obscure these electromagnetic  interactions.  However, at impact parameters large enough so that no hadronic interactions occur, the photonuclear interactions can be seen; these are  Ultra-Peripheral Collisions (UPCs).  The photon flux  is well described within the Weizs\"{a}cker-Williams formalism \cite{Baur:2001jj,Bertulani:2005ru}.   Since they come from nuclei, these photons are nearly real, with virtuality set by the nuclear radius $R_A$.  For gold, $\langle Q^{2}\rangle\sim (\hbar/R_A)^2 \sim 10^{-3}$\ (GeV/c)$^{2}$.  

Vector meson photoproduction may be modeled by the photon fluctuating to a quark-antiquark pair which then scatters from the target nucleus, emerging as a real vector meson.  A more detailed model treats the photon as a combination of Fock states: a bare photon with virtual $q\overline q$ pairs, plus higher virtual states.  This model  described the photoproduction measurements performed at HERA \cite{Bartels:2002cj} and  is also  applicable in the UPC environment.    The cross-section for UPC photoproduction can be found by convoluting the photon flux (with the constraint that there be no hadronic interactions) with the photon-nucleon cross-section.   For nuclear targets, one needs to account for the possibility of multiple dipole-target interactions, usually via a Glauber calculation.

The first calculation of  UPC photoproduction cross sections used HERA data on $\gamma p\rightarrow \rho^0 p$ as input to a classical Glauber calculation to predict the cross section with heavy ion targets  \cite{Klein:1999qj}.  It correctly predicted the $\rho^0$ photoproduction cross section at 
Relativistic Heavy Ion Collider (RHIC), at energies of 62 GeV/nucleon-pair \cite{Agakishiev:2011me}, 130 \cite{Adler:2002sc} and 200 GeV/nucleon-pair \cite{Abelev:2007nb}, and up to 2.76 TeV/nucleon-pair at the LHC \cite{Adam:2015gsa}.   A later calculation treated the $q\overline q$ pair as a dipole in a quantum Glauber calculation, which found a cross section about 50\% higher, in tension with the data \cite{Frankfurt:2002sv}.  Recently, a modification of the quantum Glauber calculation has been proposed, in which nuclear shadowing reduces the calculated $\rho^0$ cross section \cite{Frankfurt:2015cwa}.  This calculation matched the data quite well.   Other calculations include nuclear saturation mechanisms, including the color glass condensate \cite{Goncalves:2009kda,Santos:2014vwa}.   Two-photon production of $\pi^+\pi^-$ pairs also occurs, but the cross section is much smaller than for photonuclear interactions \cite{Klusek-Gawenda:2013rtu}. 

For photoproduction of $\rho^0$ mesons in gold-gold collisions at a center of mass energy of 200 GeV/nucleon-pair at RHIC, the rapidity range $|y|<0.7$ corresponds to photon-nucleon center-of-mass energies from 9 to 18 {\rm GeV}, depending on the rapidity and final state transverse momentum.    In this region,  the $\rho^{0}$ photoproduction cross section increases slowly with collision energy and the $\gamma p\rightarrow \rho^0 p$  cross section is well described by the soft-Pomeron model \cite{Crittenden:1997yz}; the $\gamma A$ cross-section is almost independent of energy \cite{Klein:1999qj}.

Because of the high photon flux, these UPC events have a high probability to be accompanied by additional photon exchanges that excite one or both of the ions into Giant Dipole Resonances (GDRs) or higher excitations.  The GDRs typically decay by emitting a single neutron, while higher resonances usually decay by emitting two or more neutrons \cite{Berman:1975tt}.   These neutrons have low momentum with respect to their parent ion, so largely retain the beam rapidity.   
For heavy nuclei, the cross section for multi-photon interactions nearly factorizes \cite{Baur:2003ar}, with the combined cross section given by an integral over impact parameter space:
\begin{equation}
\sigma(A_1A_2\rightarrow A_1^*A_2^*\rho^0) = \int d^2b\ [1-P_{\rm Had}(b)] P_1(b,A^*)P_2(b,A^*)P(b,\rho^0),
\end{equation}
where $P_{\rm Had}(b)$,  $P_1(b,A^*)$, $P_2(b,A^*)$ and $P(b,\rho^0)$ are the respective probabilities for having a hadronic interaction, exciting each of the ions and producing a $\rho^0$.  Each photon-mediated reaction occurs via independent photon exchange, so all four probabilities are tied together only through a common impact parameter \cite{Baltz:2002pp}.  The photonuclear cross sections are based on a parameterization of data \cite{Baltz:1996as}.   
A unitarization process is employed to account for the possibility of multiple photons exciting a single nucleus.  Experimentally, requiring mutual Coulomb excitation along with dipion production leads to a trigger with a higher purity, allowing more events to be collected than for the dipion state by itself.

This letter reports on the measurement of exclusive $\rho^0$ and $\omega$ meson and direct $\pi^+\pi^-$ photoproduction in UPCs between gold ions using the Solenoidal Tracker At RHIC (STAR) detector at a center-of-mass energy of 200 GeV/nucleon-pair.   The current data sample is about 100 times larger than in previous RHIC measurements   \cite{Abelev:2007nb}, allowing for much higher precision studies and two main new results.  First, the $\pi\pi$ invariant mass distribution cannot be fitted with just $\rho^0$ and direct $\pi^+\pi^-$ components; an additional contribution from photoproduction of $\omega$, with $\omega\rightarrow\pi^+\pi^-$ is required for an acceptable fit.   The second result is the observation of a detailed diffraction pattern, clearly showing the first and second minima, with a possible third.  This diffraction pattern can be used to determine the distribution of the hadronic interactions in gold nuclei.  

\section{Experimental Setup and Analysis}

This analysis uses an integrated luminosity of $1100\pm 100\ \mu$b$^{-1}$ of data collected in 2010.   Four types of STAR subsystems were used  for triggering and event reconstruction in the analysis: the Time Projection Chamber (TPC), Time of Flight system (TOF), Beam Beam Counters (BBCs), and East and West Zero Degree Calorimeters (ZDCs).

The STAR TPC \cite{Anderson:2003ur} efficiently detects charged tracks from mid-rapidity to pseudo-rapidities beyond $|\eta|=1.0$, using 45 layers of pad rows in a 2 m long cylinder.  In the 0.5 T solenoidal magnetic field, the momentum resolution is  $\Delta p/p = 0.005 + 0.004 p$ where $p$ is in GeV/c \cite{Anderson:2003ur}.   The TPC can also identify charged particles by their specific ionization  energy loss ($dE/dx$) in the TPC gas. The $dE/dx$ resolution is 8\% for a track that crosses 40 pad rows. This gives good pion/kaon/proton separation up to their respective rest masses.  The TOF surrounds the TPC, covering  the pseudorapidity region  $|\eta|<1$ \cite{Llope:2005yw}.  For this analysis, the TOF was used to reject tracks that are out of time with the beam crossing.

The other detector components were used solely for triggering.  At higher rapidities, charged particles are detected using the two BBCs, one on each side of the nominal interaction point.  Each is  formed with 18 scintillator tiles arranged around the beam pipe, covering a pseudo-rapidity window of $2<|\eta|<5$ \cite{Kiryluk:2005gg}.  The ZDCs are small hadron calorimeters installed downstream of the collision region to detect neutrons at or near beam rapidity \cite{Adler:2003sp}. 
  
The trigger \cite{Bieser:2002ah} selected  38 million events with low multiplicity in the central detector, along with one to roughly four neutrons in each ZDC, along the lines described in \cite{Abelev:2007nb}.  It required low activity in the TOF detector (at least two and  no more than six hits), no charged particles detected in the BBC detectors and finally, showers in both ZDC detectors.  The ZDC signals were required to be between 50 and 1200 ADC counts, corresponding to an energy deposition between 1/4 and about 4 beam-energy neutrons.  The one-neutron peak was centered at 198 counts, with a width ($1 \sigma$) about 55 counts, making the ZDCs almost fully efficient for single neutrons. 

The analysis selected events containing a pair of oppositely charged tracks that were consistent with originating from a single vertex, located within 50 cm longitudinally of the nominal interaction point.  The tracks were required to have at least 14 hits in the TPC (out of a possible 45), and have $dE/dx$ values within 3$\sigma$ of the expected $dE/dx$ for a pion.  Both tracks in each pair were required to have a valid hit in the TOF system to reject tracks from other beam crossings.  This requirement also limited the track acceptance largely to the region  $|\eta|<1.0$.  The 384,000 events with a $\pi^+\pi^-$ pair invariant mass in the range $0.25 < M_{\pi\pi} < 1.5$ GeV/c$^2$ were saved for further evaluation.   

The largest backgrounds for this analysis are low-multiplicity hadronic interactions (peripheral ion-ion collisions).  Other backgrounds come from other UPC reactions or from cosmic-rays accompanied by in-time mutual Coulomb excitation.   Pure electromagnetic production of  $e^+e^-$ pairs contribute less than 4\% to the $\rho^0$ peak \cite{Adler:2002sc}.   The decay $\omega\rightarrow\pi^+\pi^-\pi^0$ produces a $\pi^+\pi^-$ pair with a larger $p_T$ than for coherent photoproduction, and a 
pair invariant mass that is usually below 600 MeV.   It was a 2.7\% background in a previous analysis \cite{Abelev:2007nb}, and, due to a higher cut on the pair invariant mass, should be smaller here.  We neglect these minor backgrounds here; they are well within the overall systematic errors. 

\begin{figure}[t]
\center{\includegraphics[width=0.8\columnwidth]{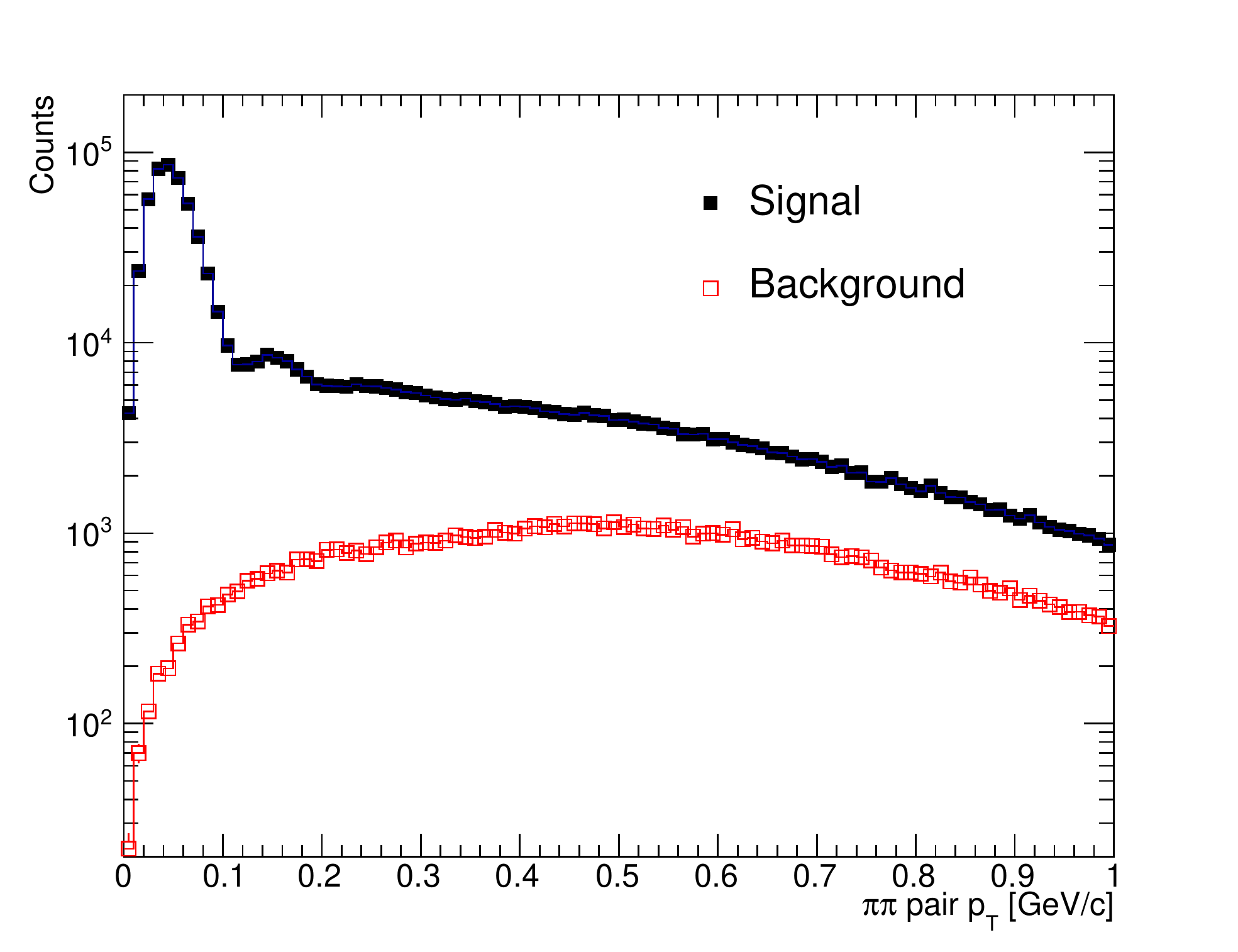}}
\caption{\label{fig:pionPairPt}The unlike-sign  (black filled squares) pion pair transverse momentum distribution. The peak below 100 MeV/c is from coherently produced $\pi^+\pi^-$ pairs. The red open squares show  the pair momentum for same-sign pion pairs. Both histograms show pairs that come from vertices with only two tracks. }
\end{figure}

The hadronic background is estimated from the like-sign pion pairs.   Figure  \ref{fig:pionPairPt} compares the transverse momentum ($p_T$) distribution of the $\pi^+\pi^-$ pairs (black histogram) with the corresponding distribution for like-sign pairs (red histogram) in  vertices recorded
 with only two tracks.  The signal distribution has a prominent peak for $p_T <  100$  MeV/c, from coherent photoproduction of pion pairs from the gold nucleus.  
 
\begin{figure}[t]
\includegraphics[width=0.95\columnwidth]{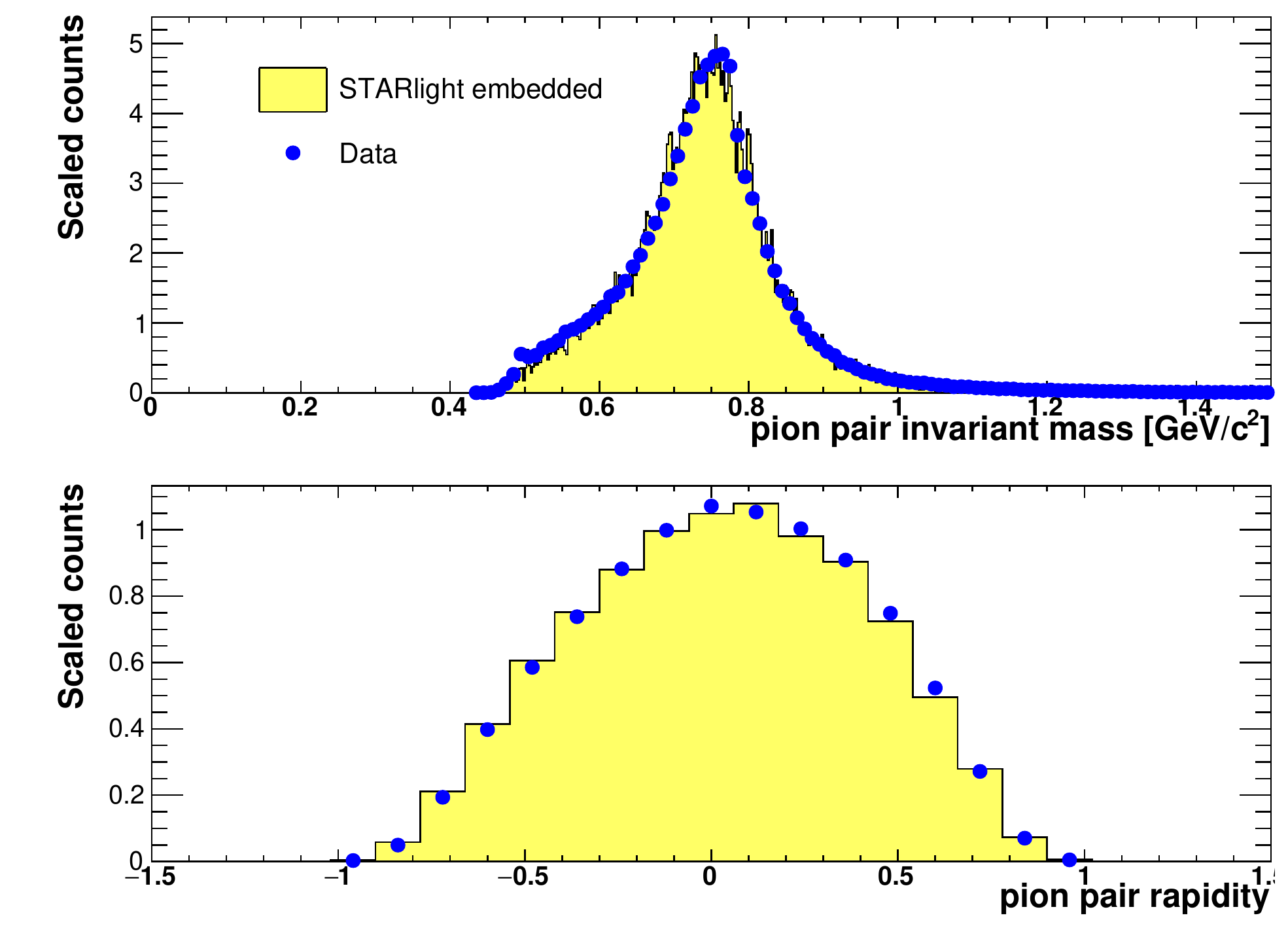}
\caption{\label{fig:validation}Plots comparing data and the simulations used for efficiency determination, after all cuts. Comparison of uncorrected data (blue points) with embedded simulated $\rho^0$ and direct $\pi^+\pi^-$ events (yellow histogram).  The simulated UPCs were run through a GEANT simulation of the detector, embedded in randomly triggered (zero-bias) events, and subject to the same reconstruction programs as the data.}
\end{figure}

The reconstructed events are corrected for acceptance and detection efficiency using a detailed simulation of the STAR detector.  A mix of $\rho^0$ mesons and non-resonant $\pi^+\pi^-$ events are generated using the STARlight Monte Carlo  \cite{Klein:2016yzr, Klein:1999qj} which reproduces the kinematics of the processes, including the mass and rapidity distributions.  The generated events are passed through a complete GEANT \cite{Brun:1994aa} simulation of the detector  and then embedded in `zero bias' STAR events.
Zero-bias events are data from randomly selected beam crossings.  This embedding procedure accurately accounts for the detector noise and backgrounds, including overlapping events recorded in the STAR TPC during its sizable active time windows.   As Fig. \ref{fig:validation} shows, the agreement between the Monte Carlo and data is very good.   The agreement in both pair mass and rapidity and other kinematic distributions (not shown) gives us confidence that the Monte Carlo will correctly predict the experimental acceptance.

The event reconstruction efficiency depends only weakly on the pair mass and pair $p_T$, but depends fairly strongly on rapidity.   The rapidity dependence  has a bell shape with a maximum of 13\% efficiency at $y\approx 0.1$.  It is slightly asymmetric because of  inefficiencies in one of the TPC East (rapidity $<0$) sectors. 
One uncertainty in the reconstruction efficiency stems from uncertainties in the actual (`as-built') positions of the TOF slats, which may not be completely accurately reflected in the simulations. While this uncertainty may affect the measured $d\sigma/dy$, particularly at large rapidity, it does not significantly affect the pair $p_T$ or mass acceptance uncertainties.

\begin{figure}[t]
\center{\includegraphics[width=0.75\columnwidth]{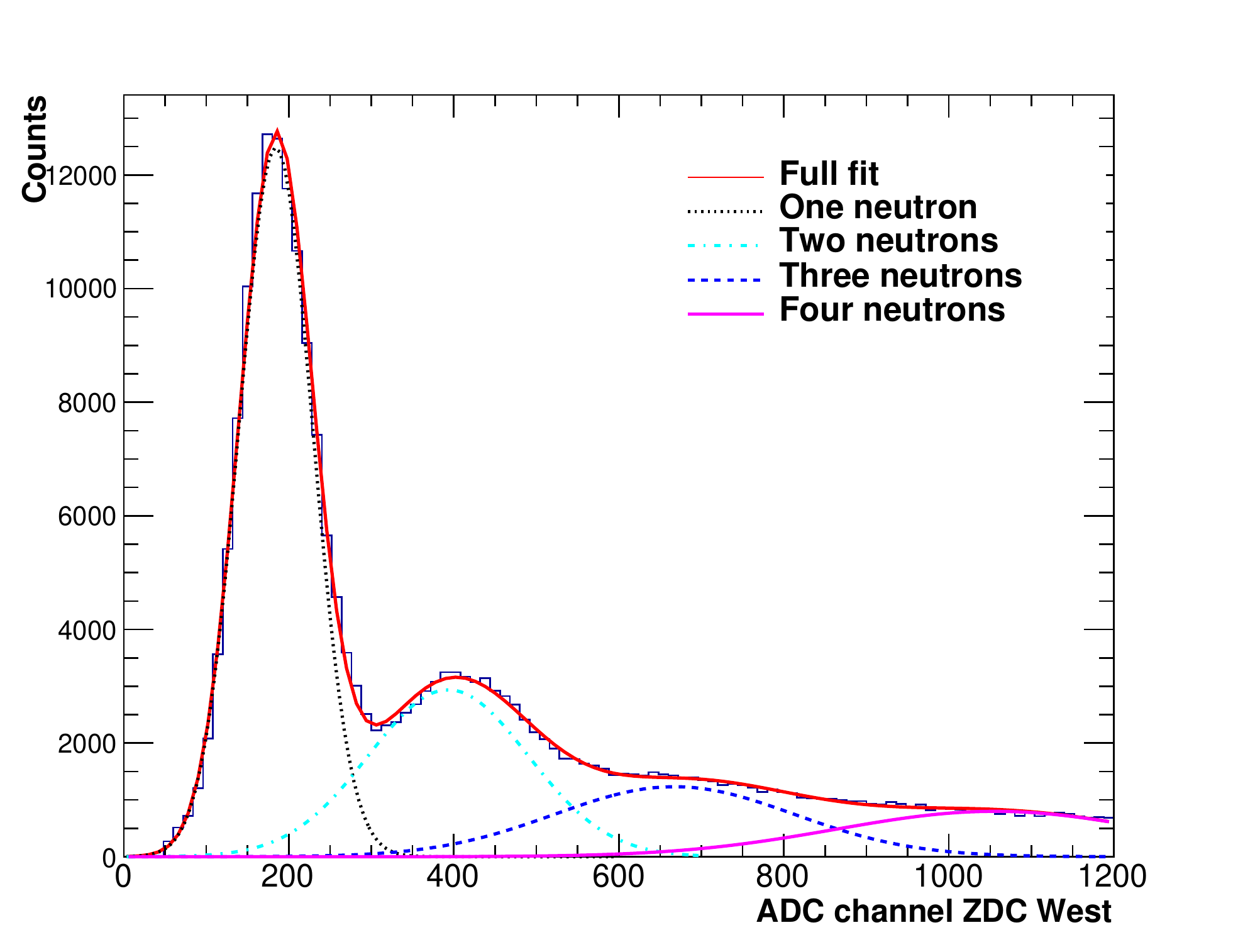}}
\caption{\label{fig:ZDCfit}The shower energy in the West ZDC from neutrons produced by mutual
dissociation is shown as a distribution of ADC channels. These events have a single
neutron detected on the East ZDC. The peaks corresponding to 1 to 4 neutrons are fitted with Gaussian distributions with standard deviations that grow as $n\sigma$ with $n$ the number of neutrons and $\sigma$ the standard deviation of the one neutron  Gaussian. 
The red curve is the sum of all Gaussians which are also displayed individually.}
\end{figure}

This analysis considers two classes of nuclear breakup: single neutrons (1n), associated with Giant Dipole Resonances, or more than one neutron (Xn), from a broad range of photonuclear interactions.  Figure \ref{fig:ZDCfit} shows the ADC distribution from the West ZDC for events that satisfy a cut which selects events with a single neutron in the East ZDC and a photoproduced \myro with $|y|<1$ and $p_T < 100$ MeV/c.      Table \ref{tab:neutronCS} shows the cross-sections for coherent $\rho^0$ photoproduction accompanied by different numbers of neutrons.  There is some non-linearity in the system.  The cross sections are determined by applying a window to one ZDC spectrum and fitting the neutron spectrum in the other, and then reversing the procedure.  The fits included events with one, two, three, or four neutrons in each ZDC.   The one and two neutron peaks are very clear, but the higher peaks are less obvious. The two results are averaged, and the difference is used as an estimate of the systematic error.  Statistical errors are $< 1\%$ and are not listed.  Systematic errors arising from the event-selection cuts were added in quadrature to the quadrature sum of the relevant common uncertainties listed in Tab. \ref{tab:common} (17\%).

The limited ZDC window led to a relatively high yield of photoproduced $\rho^0$ per trigger, but the cost was that it did not cover the full neutron number spectrum.   So, we used the \oneN events to normalize the XnXn cross section, based on the STARlight \cite{Klein:2016yzr} calculation of the cross section ratio.  We find the ratio of triggered events to those with single neutrons in each ZDC, using the fit results in Tab. \ref{tab:neutronCS}, and use the STARlight ratio of \xN to \oneN events to normalize the overall cross section scale.

The cross sections in Tab. \ref{tab:neutronCS} decrease slowly with increasing total neutron number. The summed cross section for 
 $2\rm{n}1\rm{n} + 1\rm{n}2\rm{n}$ ({\it i.e.} the two combinations with 1 neutron in one direction),  is 83\% of the \oneN cross section.  This fraction is larger than is seen for mutual Coulomb dissociation, where one calculation has the  ($2\rm{n}1\rm{n} + 1\rm{n}2\rm{n}):1\rm{n}1\rm{n}$ ratio around 0.6 \cite{Pshenichnov:2001qd} and another finds a ratio around 0.4, albeit at a slightly lower beam energy \cite{Klusek-Gawenda:2013ema}.  Some of this difference is because the requirement of $\rho^0$ photoproduction selects events with smaller impact parameters, where the photon spectrum is harder \cite{Baur:2003ar}.  

\begin{table}
\begin{tabular} {l|lll}
East & & West ZDC & \\
 ZDC           &  1n                 & 2n                 & 3n             \\
 \hline
1n     & $1.38\pm0.24$  mb    & $0.57\pm0.11$ mb      & $0.39\pm0.07$ mb      \\
2n     & $0.57\pm0.11$   mb   & $0.23\pm0.04$  mb     & $0.18\pm0.03$  mb     \\
3n     & $0.40\pm0.07$   mb   & $0.19\pm0.03$   mb    & $0.15\pm0.03$   mb    \\

\end{tabular}
\caption{Mutual dissociation cross sections for events with exclusive coherent \myro photoproduction, broken down by the number of neutrons in the East (rows) and West (columns) ZDCs.  
\label{tab:neutronCS}
}
\end{table}

\section{The $\pi^+\pi^-$ Mass Spectrum}

\begin{figure}[t]
\includegraphics[width=\columnwidth]{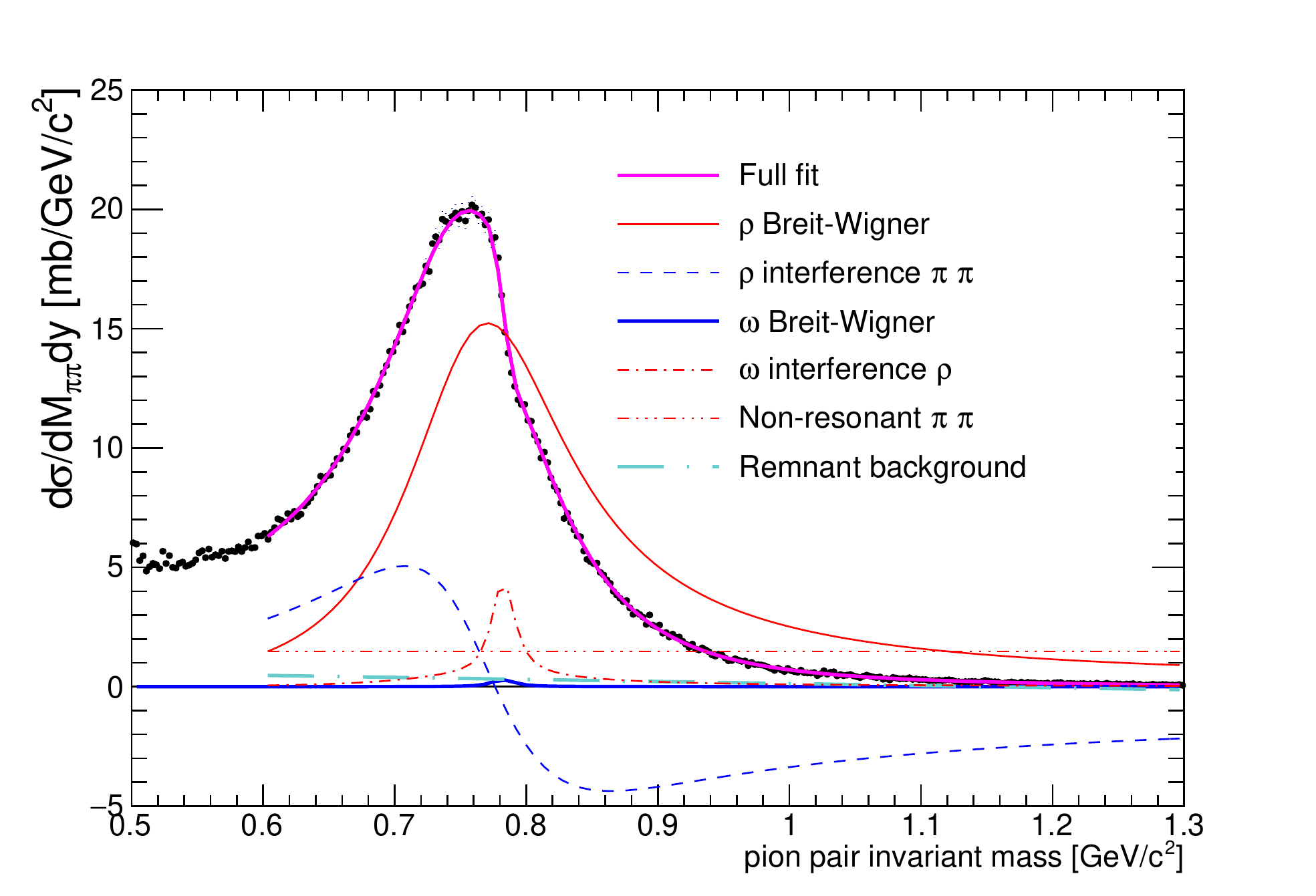}
\caption{\label{fig:invMass}The $\pi^{+} \pi^{-}$ invariant mass distribution for all selected $\pi\pi$ candidates with $p_T < 100$ MeV/c.  The black markers show the data (in 2.5 MeV/c$^2$ bins).  The magenta curve is the modified S\"{o}ding fit to the data in the range $0.6< M_{\pi\pi}<1.3 $ GeV/c$^2$.   Also shown are the 
 \myro Breit-Wigner component of the fit (brown curve),  constant non-resonant pion pair component  (brown-dashed curve),  interference between non-resonant pion pairs and the \myro (blue-dashed curve),   Breit-Wigner distribution for the $\omega$ mesons (blue solid curve), interference between \myro and $\omega$  (red-dashed curve), and a small  contribution from the remnant background, fit by a linear polynomial (cyan-dashed curve).}
\end{figure}

 Figure \ref{fig:invMass} shows the efficiency-corrected, like-sign-pair (background) subtracted invariant mass of the  pion pairs with $p_T < 100$ MeV/c.   Events with dipion mass $M_{\pi\pi}> 600$ MeV/c$^2$ were initially fitted with a modified  S\"{o}ding parametrization 
\cite{Soding:1965nh} which included a relativistic Breit-Wigner resonance for the $\rho^0$ plus a flat direct $\pi^+\pi^-$ continuum.   This 2-component model was a poor fit to the data, 
so an additional relativistic Breit-Wigner component was added, to account for $\omega$ photoproduction, followed by its decay $\omega\rightarrow\pi^+\pi^-$. This leads to the following fit function:

\begin{widetext}
\begin{equation}
 \frac{d\sigma}{dM_{\pi^{+}\pi^{-}}}\! \propto\! \left | A_{\rho}\frac{\sqrt{M_{\pi\pi}M_{\rho}\Gamma_{\rho}}}{M_{\pi\pi}^{2}-M_{\rho}^{2}+iM_{\rho}\Gamma_{\rho}}+B_{\pi\pi}+C_{\omega}e^{i\phi_{\omega}}\frac{\sqrt{M_{\pi\pi}M_{\omega}\Gamma_{\omega\rightarrow\pi\pi}}}{M_{\pi\pi}^{2}-M_{\omega}^{2}+iM_{\omega}\Gamma_{\omega}} \right |^{2} + f_{p}
\label{massFitFunction}
\end{equation}
\end{widetext}\
where $A_\rho$ is the $\rho$ amplitude, $B_{\pi\pi}$ is the amplitude for the direct pions, $C_\omega$ is the amplitude for the $\omega$, and $f_p$ is a linear polynomial that accounts for the remaining background.  The momentum-dependent widths in Eqs. (3) and (4) below are motivated by the forms proposed in 
Ref. \cite{Alvensleben:1971hz}, where $\Gamma_0$ is the pole width for each meson. Several variations of the dipion mass dependence for the $\omega$ width were tried, but none were significantly different from a constant, reflecting the fact that the $\omega$ width is small, and the width does not change significantly in that mass range.  The  momentum-dependent widths are taken to be
\begin{equation}
\Gamma_{\rho} = \Gamma_{0} \frac{M_{\rho}}{M_{\pi\pi}}\left(\frac{M_{\pi\pi}^{2}-4m_{\pi}^{2}}{M_{\rho}^{2}-4m_{\pi}^{2}}\right)^{3/2}
\end{equation}
and 
\begin{equation}
\Gamma_{\omega} = \Gamma_{0} \frac{M_{\omega}}{M_{\pi\pi}}\left(\frac{M_{\pi\pi}^{2}-9m_{\pi}^{2}}{M_{\omega}^{2}-9m_{\pi}^{2}}\right)^{n},
\end{equation}
where $\Gamma_{0}$ is the  pole width for each meson.  For the $\omega$, the $9m_\pi^2$ term reflects the fact that the $\omega$ decay is dominated by the three-pion channel, $n=3/2$ for a quasi-two-body decay \cite{Alvensleben:1971hz} and $n=4$ for a free-space three-body decay \cite{Kuraev:1995hc,Akhmetshin:2000ca}. We have tested $\Gamma$ as constant, and the  $n=3/2$ and $n=4$ boundary cases. All three fits result in  negligible difference due to the narrow width of $\omega$ decay, and we choose a default $\Gamma$ with $n=3/2$ for all the fits shown in the figures and extracted values.  The branching ratio for $\omega\rightarrow\pi^+\pi^-$ is small, so we use
\begin{equation}
\Gamma_{\omega\rightarrow\pi\pi} = {\rm Br}({\omega\rightarrow\pi\pi}) \Gamma_0 
\frac{M_\omega}{M_{\pi\pi}}
\left(\frac{M_{\pi\pi}^{2}-4m_{\pi}^{2}}{M_{\omega}^{2}-4m_{\pi}^{2}}\right)^{3/2}
\end{equation}
with ${\rm Br}(\omega\rightarrow\pi\pi) = 0.0153^{+0.0011} _{-0.0013}$ \cite{Agashe:2014kda}. 

In Eq. \ref{massFitFunction}, $f_{p}$ is a linear function that describes the remaining remnant background.  The masses and widths of the $\rho^0$ and $\omega$ were allowed to float, giving a total of ten parameters: two masses, two widths, three amplitudes, the phase of the $\omega$ meson, and two parameters for the background.

Figure \ref{fig:invMass}, shows the data, the full fit function, and most of the components, while Tab. \ref{tab:fit} shows the fit results.  The $\rho^0$ and $\omega$ masses and the $\rho^0$ width are in  good agreement with their Particle Data Group values \cite{Agashe:2014kda}.  The $\omega$ is considerably wider than the standard value, because it is broadened by the detector resolution, which is comparable to the $\omega$ width.  The fit $\chi^2/DOF=255/270$ shows that the data and model are consistent in the fit region.

The ratio of direct $\pi^+\pi^-$ to $\rho^0$  amplitudes, $|B/A|=0.79 \pm 0.01~ (stat.) \pm 0.08~ (syst.)$ (GeV/c$^2)^{-1/2}$,  agrees within the $1 \sigma$ uncertainty with the value reported in the previous STAR publication \cite{Abelev:2007nb}: $0.89\pm0.08~(stat.)\pm0.09~(syst.)$ (GeV/c$^2)^{-1/2}$.  At 2.76 TeV/nucleon-pair, the ALICE collaboration measured a smaller ratio, $|B/A| =  0.50\pm0.04~(stat.)^{+0.10}_{-0.04}~(syst.)$ (GeV/c$^2)^{-1/2}$ \cite{Adam:2015gsa}.

The measured ratio of $\omega$ to $\rho^0$ amplitude was $C/A = 0.36 \pm 0.03~ (stat.) \pm 0.04~ (syst.)$.  The $\omega$ amplitude is small, but is clearly visible through its interference with the $\rho^0$ which produces a small kink in the spectrum near 800 MeV/c$^2$.    The $\omega$ amplitude agrees with a prediction from STARlight \cite{Klein:2016yzr}, $C/A =0.32$,  which uses the $\gamma p\rightarrow \omega p$ cross section and a classical Glauber calculation.

The only previous measurement of $\rho^0$-$\omega$ interference in the $\pi^+\pi^-$ channel was made by a DESY-MIT group, using $5-7$ GeV photon beams \cite{Alvensleben:1971hz}.  That fit used a similar but not identical fit function.   Neglecting some differences in the treatment of the $\omega$ width,  that result was, in our terminology,  $|C/A|=0.36\pm 0.04$.  In the terminology of Ref.  \cite{Alvensleben:1971hz} $|C/A|=\zeta\sqrt{M_\rho\Gamma_\rho/M_\omega\Gamma_\omega}/\sqrt{(Br(\omega\rightarrow\pi\pi)}$, where $\zeta$ is their $\omega$ amplitude.

Our fit finds a non-zero $\omega$ phase angle, $\phi_\omega = 1.46\pm0.11(stat.)\pm 0.07(syst.)$.   The systematic error was estimated from fits using slightly different fit functions.  
This phase angle result is a bit lower than, but consistent within experimental uncertainties with the DESY-MIT measurement of $1.68 \pm 0.26$.   The DESY-MIT experiment used much lower energy photons, in a regime where $\omega$ production proceeds via both meson and Pomeron exchange. 
This shows that the $\rho$ and $\omega$ phases are either relatively constant, or change in tandem over a fairly wide range of photon energy.  Other experiments have studied $\rho^0-\omega$ interference using photoproduction to the $e^+e^-$ final state (where the $\omega$ is more visible but the branching ratios are much smaller), or via the reaction $e^+e^-\rightarrow \pi^+\pi^-$, and found similar phase angles \cite{Lemke:1972mn,Langacker:1979cy}. 

\begin{table}
\begin{tabular} {lrc}
Fit Parameter               & value                & units                        \\
 \hline
$M_{\rho}$                  & $0.7762 \pm 0.0006 $ & {\rm GeV}/c$^2$                    \\
$\Gamma_{\rho}$         & $0.156 \pm 0.001 $ &  {\rm GeV}/c$^2$                   \\
$A_{\rho}$                  & $1.538  \pm 0.005  $ &                               \\
$B_{\pi\pi}$                & $-1.21 \pm 0.01  $ & $({\rm GeV/c}^2)^{-1/2}$     \\
$C_{\omega}$                & $0.55 \pm 0.04  $ &                       \\
$M_{\omega}$                & $0.7824 \pm 0.0008 $ & {\rm GeV}/c$^2$             \\
$\Gamma_{\omega}$           & $0.017 \pm 0.002 $ & {\rm GeV}/c$^2$             \\
$\phi_{\omega}$             & $1.46   \pm 0.11   $ &  radians              \\      
$f_{p}~ p_{0}$              & $0.99  \pm 0.07  $ &   $({\rm GeV}/c^2)^{-1}$                    \\
$f_{p}~ p_{1}$              & $-0.86 \pm 0.06   $ &   $({\rm GeV}/c^2)^{-2}$                     \\
\end{tabular}
\caption{The results of fitting Eq. \ref{massFitFunction} to the data. The  parameters $p_0$ and $p_1$ are for the polynomial background. 
\label{tab:fit}
}
\end{table}

An alternate fit was performed, where $B_{\pi\pi}$ was multiplied by a mass dependent term, ($M_{\rho}/M_{\pi\pi})^2 [(M_{\pi\pi}^2/4-m_{\pi}^2)/(M_\rho^2/4-m_{\pi}^2)]^{3/4}$ \cite{Pumplin:1970kp} to account for the possibility that the continuum $\pi\pi$ pairs do not completely interfere with the $\rho^0$ or $\omega$.   This fit produced similar results, with a comparable $\chi^2/DOF$. 

To study the photon energy dependence of the amplitude ratios, we performed the fit in five bins of rapidity: $y<-0.35$, $-0.35 < y < -0.15$, $-0.15 < y <0.15$, $0.15 < y <0.35$,  and $y>0.35$.   These bins were chosen so that each of the three $|y|$ ranges included about 100,000 pion pairs.  The amplitudes should be symmetric around $y=0$; pairing by $|y|$ provides a check on rapidity-dependent systematic errors.  To ensure the fits were stable, the values of $M_{\omega}$ and $\Gamma_{\omega}$ were fixed to the values extracted from the fit to the rapidity-integrated pion pair mass distribution.

In the lab frame, at low $p_T$,  the rapidity is related to photon energy $k$ by 
\begin{equation}
k=M_{\pi\pi}/2\exp{(\pm y)}.
\label{eq:ktorap}
\end{equation}
The $\pm$ sign reflects the two-fold ambiguity as to which nucleus emitted the photon.  Table \ref{tab:raptok} gives the lab-frame photon energies and the $\gamma N$ center-of-mass energies for the two solutions to Eq. \ref{eq:ktorap} for the centers of the rapidity bins when $M_{\pi\pi}=M_\rho$. The photon flux drops rapidly with increasing energy, so away from $y=0$, the cross section is dominated by the lower photon energy; the relative fractions scale roughly as the ratio of the lab-frame photon energies. 
\begin{table*}
\begin{tabular} {l|cc}
Rapidity   & Photon Energy (lab frame) & $\gamma N$ center-of-mass energy \\
              &  (MeV)                                & (GeV) \\
 \hline
0   & 380  &   12.4   \\
0.15 & 327 &  11.5    \\
        & 441  & 13.4    \\
0.4  &  255   & 10.2      \\
        & 488    &  14.1    \\
0.63 &  202    &  9.1        \\
        &  713    &  17.0 
\end{tabular}
\caption{Photon energy (lab frame) and $\gamma N$ center-of-mass energy for different rapidities.  There are two rows per rapidity, one for the higher energy photon solution, and one for the lower one.
} 
\label{tab:raptok}
\end{table*}

\begin{figure}[tbh]
\center{\includegraphics[width=0.6\columnwidth]{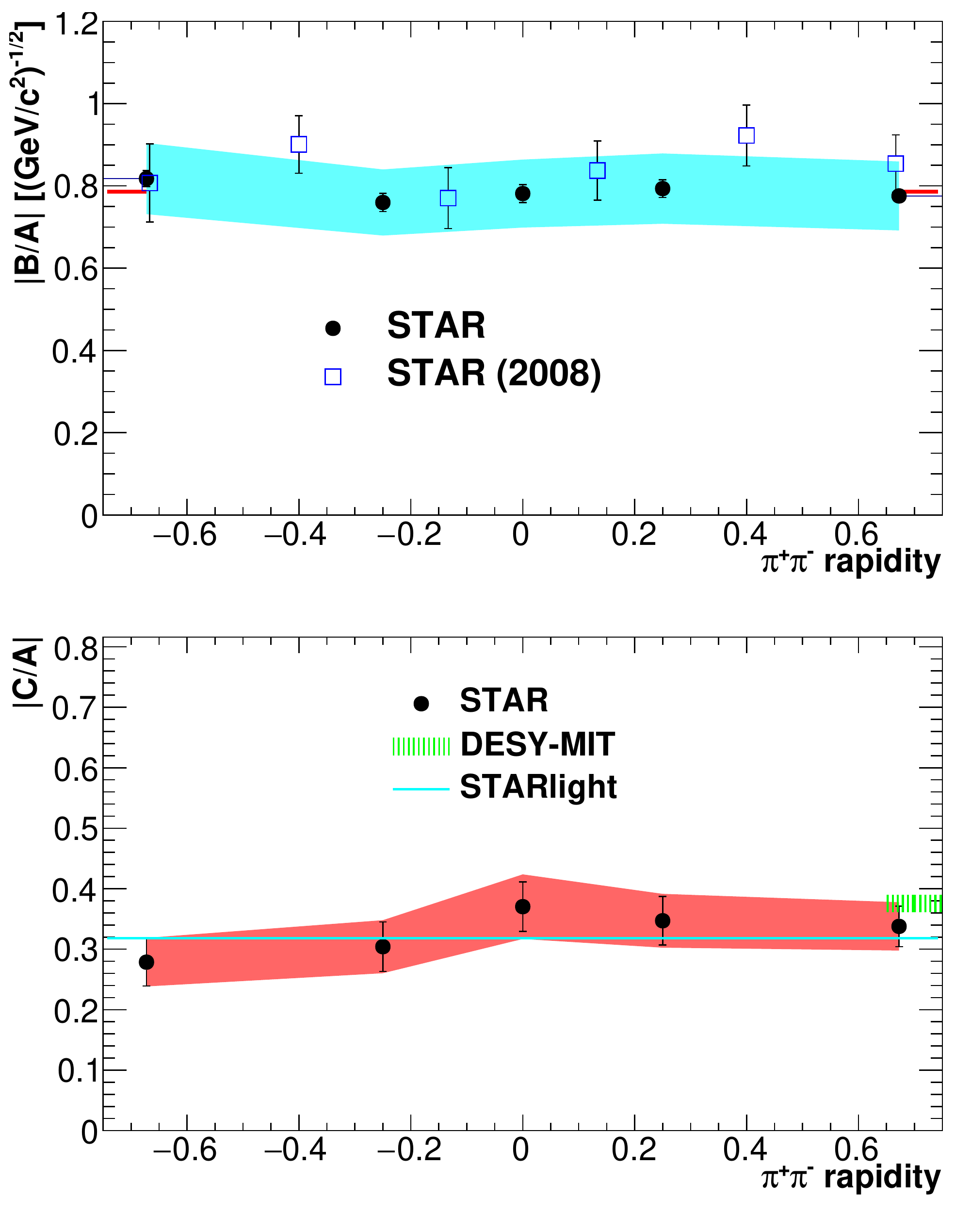}}
\caption{(Top) The ratio $|B/A|$ of amplitudes of non-resonant \pionpair and \myro mesons. The black points (with shaded blue systematic error band) are from the current analysis, while the previous STAR results are shown with open blue squares.  The red line shows the rapidity-averaged result.  In the bottom panel, the black points show the ratio $|C/A|$ of the $\omega$ to \myro amplitude.  The red band shows the systematic errors, while the horizontal blue line shows the STARlight prediction with the most recent branching ratio for $\omega\rightarrow\pi^{+}\pi^{-}$ decay \cite{Agashe:2014kda}.
The green dashed band shows the DESY-MIT result for $|C/A|$ \cite{Alvensleben:1971hz}.  Their result was at much lower photon energies, equivalent to a large effective rapidity.  For the lower energy photon solution of the two-fold ambiguity, the effective rapidity is about $-2.5$.
 \label{fig:BoverA}
}
\end{figure}

Figure \ref{fig:BoverA} shows the direct $\pi^+\pi^-$ to $\rho^0$  ($|B/A|$) and $\omega$ to $\rho^0$ ($C/A$) ratios in the five rapidity bins.  Both $|B/A|$ and $C/A$ are unchanged as rapidity varies, showing  that these ratios do not have a large dependence on the photon energy.    Also shown are the STARlight predictions and, for $C/A$, the DESY-MIT result.  The DESY-MIT result is at a much lower beam energy which would correspond to an effective rapidity of $-2.5$ with the lower photon energy solution of Eq. \ref{eq:ktorap}.

To determine the $\rho^0$ cross section as a function of rapidity, we integrate the $\rho^0$ Breit-Wigner function over the mass range from $2M_{\pi}$ to $M_\rho+5\Gamma_{\rho}$.  

\begin{figure}
\includegraphics[width=0.7\columnwidth]{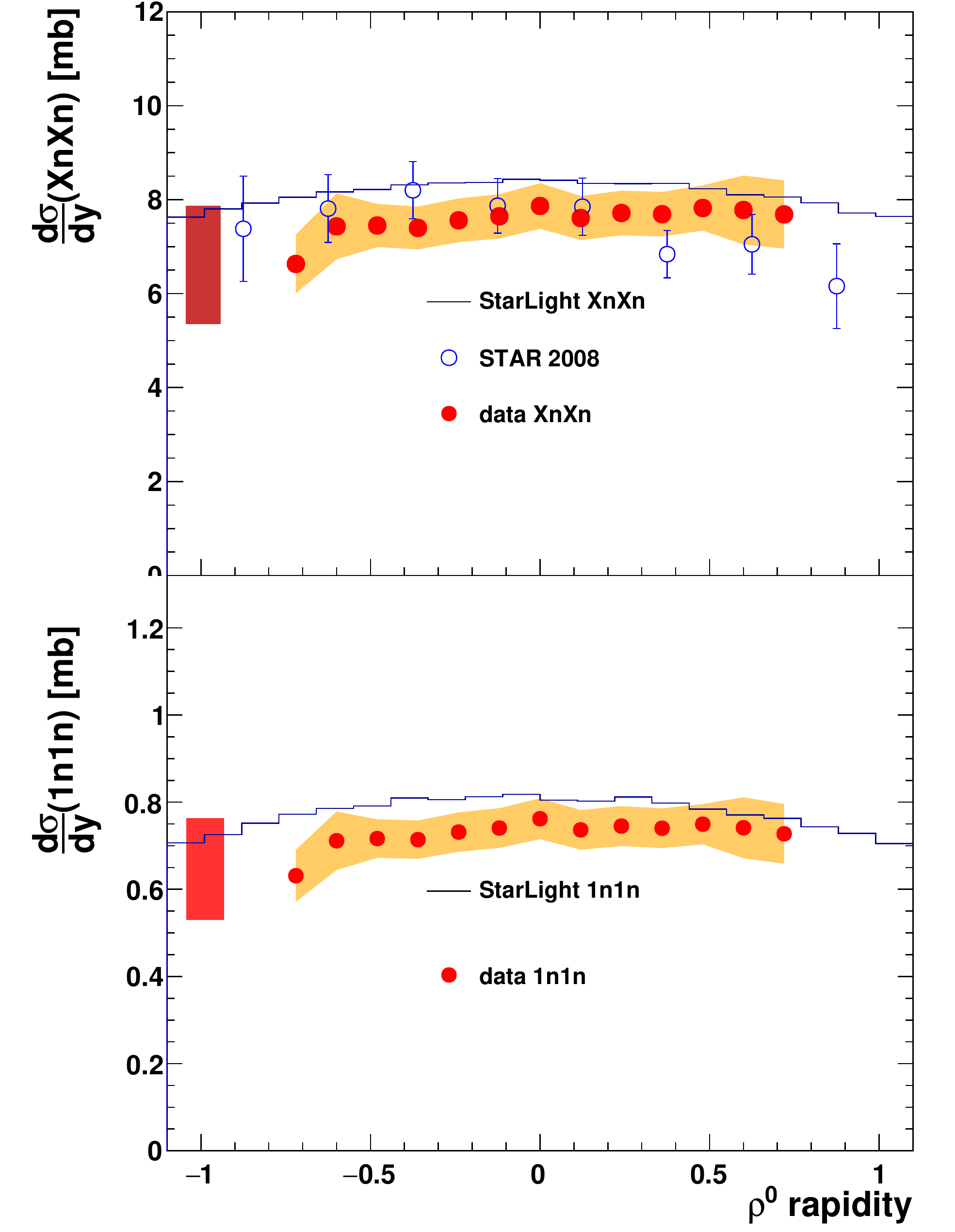}
\caption{
\label{fig:rapidityRho}
$d\sigma/dy$ for exclusively photoproduced $\rho^0$ mesons  in (top) XnXn events and (bottom) 1n1n events.  The data are shown with red markers. The statistical errors are smaller than the symbols, the orange band shows the quadrature sum of the point-to-point systematic uncertainties. The red boxes at $y\approx -0.9$ show the quadrature sum of the common systematic uncertainties. The black histograms are the STARlight calculation for \myro mesons with mutual dissociation.   The blue markers in the top panel show the previous STAR measurement \cite{Abelev:2007nb}.
}
\end{figure}

Figure \ref{fig:rapidityRho} shows the acceptance corrected $d\sigma/dy$ for $\rho^0 $.  The asymmetry between positive and negative rapidity gives a measure of the rapidity-dependent systematic uncertainties in the cross section.  This is likely due to asymmetries in the as-built longitudinal position of the TOF counters. The magnitude of this uncertainty grows slowly with increasing rapidity, reaching 4\% at $y=0.7$.  Since the actual lengths of the TOF slats are known, this uncertainty does not apply for rapidity-integrated measurements.

The systematic uncertainties in these measurements fall into two classes, either an overall scale for the cross section, or uncertainties that vary point-to-point.  The former is usually dominant.    

The uncertainty in the integrated luminosity is 10\%.  As with previous measurements \cite{Abelev:2007nb}, this uncertainty is mainly driven by the fraction of the total Au+Au cross section accessible with the trigger used to collect this data.  The selection of the number of neutrons produced in mutual electromagnetic dissociation depends on the response of the ZDC calorimeters.  We allocate a 5\% uncertainty to this neutron counting due to small non-linearities in the calorimeters and overlaps between one and many neutron distributions. We assign a 7\% uncertainty due to modeling of the TOF system in the simulation, based on studies of the TOF response in more central collisions. The track reconstruction efficiency for the STAR TPC has a 3\% per track uncertainty \cite{Anderson:2003ur} (6\% for two tracks) while the efficiency of the vertex finder is known within a 5\% uncertainty,  driven by the effect of backgrounds.
The uncertainty in how often the BBC detectors will veto good UPC events is due to fluctuating backgrounds. Even with use of embedding techniques, we estimate that these veto conditions 
introduce a 2\% uncertainty to the results.   

The same-sign pion pair distributions are the best estimators for the hadronic backgrounds for these two-track events. The background subtraction was done at the level of raw histograms and also after a fit to the background to eliminate statistical fluctuations. These two procedures lead to final results that agree within 1.5\%.

The scaling from the rapidity distribution extracted from 1n1n events to the previously measured XnXn distribution uses a correction extracted from the event generator STARlight.  There is a 6\% XnXn cross-section uncertainty from the uncertainty in the neutron data used  as input to STARlight.  This uncertainty is squared because we detect neutrons in both beams, but applies only to the XnXn results.

Table \ref{tab:common} summarizes these common systematic uncertainties. They are summed in quadrature to find the 18.2\%  overall common uncertainty.  This uncertainty is a bit higher than in our comparable previous publication \cite{Abelev:2007nb}, largely because of additional uncertainties associated with the pileup and the more complex trigger that is required to deal with the higher luminosities. 
\begin{table}
\begin{tabular} {lcccc}
Name                           & Value   & Comment                                   \\
 \hline
Luminosity                       & 10.0\%    &                                          \\
ZDC                              & 5.0\%     &  ADC ch. to num. neutrons                \\
TOF geometry modeling            & 7.0\%     &                                          \\
TPC tracking efficiency          & 6.0\%     & 3.0\% per track \cite{Anderson:2003ur}     \\
Vertex Finder efficiency         & 5.0\%     & Background driven                        \\
BBC veto in trigger              & 2.0\%     & Background driven                        \\
Efficiency determination        & 7.0\%     &              \\
Conversion from $\pi^+\pi^-$ pairs to $\rho^0$ yield & 2.2\% & Varying mass fit range \\
Background subtraction           & 1.5\%    &                                          \\
STARlight model                  & 6.0\%     &  only for XnXn results                   \\
\hline
Quadrature Sum                   & 18.2\%     &                                          \\
\end{tabular}
\caption{The common systematic uncertainties present in the rapidity distribution in Fig. \ref{fig:rapidityRho} and the $-t$ distributions in Figs. \ref{fig:fitIncoherent} and \ref{fig:dataDiffraction}.  These uncertainties are given as percentage of the measured quantities.} 
\label{tab:common}
\end{table}

The main point-to-point systematic uncertainties in the rapidity and $p_T$ distributions come from the track selection and particle identification.   The systematic uncertainties were evaluated by varying the track quality cuts and PID cuts around their central value in both the data and simulation, and seeing how the final result varies.  Table \ref{tab:tableRap} lists the point-to-point uncertainties in the rapidity distribution while Tab. \ref{tab:table2} lists the point-to-point uncertainties  for the $p_{T}$ distribution. 

\begin{table*}
\begin{tabular} {l|cccc}
Rapidity   & PID cut   & Fit to eff.& Number of track hits & TOF asymmetry   \\
 \hline
-0.7 $-$\ -0.5  & 8.\%      &   0.25\%   &     0.2\%            & 5\%    \\
-0.5 $-$  0.0    & 5.\%      &   0.25\%   &     0.05\%           & 3.6\%  \\
\ 0.0 $-$  0.5     & 5.\%      &   0.25\%   &     0.05\%           & 3.6\%  \\
\ 0.5 $-$ 0.7    & 8.\%      &   0.25\%   &     0.2\%            & 5\%    \\
\end{tabular}
\caption{Point-to-point systematic uncertainties on $d\sigma/dy$ (Fig. \ref{fig:rapidityRho}), as a percentage of the measured cross section in four rapidity ranges. PID cut refers to uncertainty in the efficiency for $\pi$ identification via the truncated $dE/dx$ \cite{Xu:2008th}.  Those cuts were varied simultaneously in the data and simulation to determine the  uncertainty in particle identification efficiency.  
The fit to efficiency is the uncertainty in the efficiency parameterization, while the number of track hits is the minimum number of points used for fitting the track.  The TOF asymmetry is the uncertainty due to the positions of the TOF slats. 
} 
\label{tab:tableRap}
\end{table*}

The ALICE collaboration has studied dipion photoproduction, in lead-lead collisions at the Large Hadron Collider (LHC) \cite{Adam:2015gsa}.   They fit their dipion mass distribution in the range from 0.6 to 1.5 GeV/c$^2$ to a function like Eq. \ref{massFitFunction}, but without the $\omega$ component, finding masses and widths consistent with the standard values.   Their cross-section values were about 10\% above the STARlight prediction. 

\begin{table}
\begin{tabular} {l|cccc}

$-t\  [({\rm GeV}/c)^{2}]$ & track sel. & pion PID & Incoh. comp. sub.   \\
 \hline
0.00 $-$ 0.02                 & 0.2\%      &   8\%    &        0.5\%            \\
0.02 $-$ 0.04              & 0.2\%      &   8\%    &        3.0\%            \\
0.04 $-$ 0.10               & 0.2\%      &   8\%    &        8.5\%            \\
\end{tabular}
\caption{Point-to-point systematic uncertainties for the $-t$ distribution shown in Fig.  \ref{fig:dataDiffraction}, as a percentage of the measured cross section in three $-t$ ranges. The PID and track selection uncertainties are described in the text. The uncertainty in the incoherent component subtraction was estimated by selecting the largest relative deviation from the default value and cross sections extracted by changing the value of the fit parameters by one standard deviation while the other parameters remain at the default fit value.} 
\label{tab:table2}
\end{table}
\section{Measurement of $d\sigma/dt$}
 
\begin{figure}
\includegraphics[width=0.8\columnwidth]{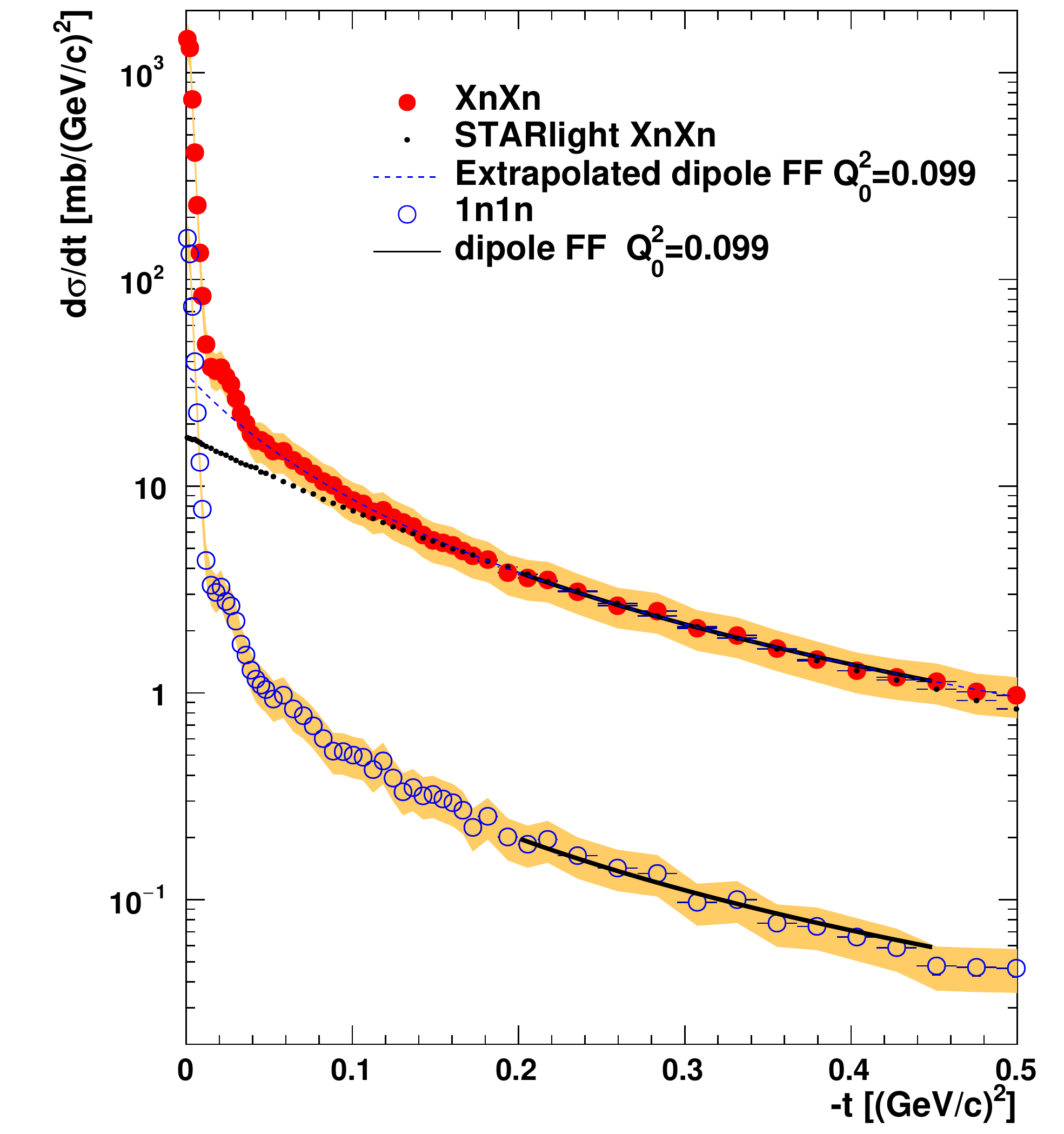}
\caption{\label{fig:fitIncoherent}The  -$t$ distribution for exclusive \myro mesons  in events with 1n1n mutual dissociation (open blue circles) and XnXn (filled red circles). the statistical errors are smaller than the points, and the colored bands show the total systematic uncertainties.   The dipole fits are shown with solid black lines. 
For XnXn, the dipole form factors are shown extrapolated to low $|t|$ (dotted black line line), along with the STARlight prediction for the incoherent contribution (dashed blue line).
 }
\end{figure}

Figure \ref{fig:fitIncoherent} shows the efficiency-corrected differential cross section $d\sigma/dt$ for \myro mesons
within the measured range $|y|<1$,  after like-sign background subtraction. 
The Mandelstam variable $t$ is expressed as $t = t_\parallel + t_\perp$ 
with $t_\parallel = -M_{\rho}^{2}/(\gamma^{2}e^{\pm y})$ and $t_\perp = -(p^{pair}_{T})^{2}$.  Here, $\gamma$ is the Lorentz boost of the ions.  At RHIC energies, $t_\parallel$ is almost negligible.   The cross section $d\sigma/dt$ for \myro mesons is obtained by scaling the total dipion cross-section by a factor of 0.75.  This factor was extracted from comparisons between the number of pion pairs with invariant masses ranging from 500 MeV/c$^2$ to 1.5 GeV/c$^2$ and the integral of the \myro Breit-Wigner function extracted from fits in rapidity and $-t$ bins. In all comparisons, the integrals are performed from $2M_{\pi}$ to $M_\rho+5\Gamma_{\rho}$.

We separate the \myro $t$-spectrum into coherent and incoherent components based on the shape of the distribution in Fig. \ref{fig:fitIncoherent}.  Because of the ZDC requirement in the trigger, and the presence of Coulomb excitation, we cannot use the presence of neutrons from nuclear breakup as an event-by-event signature of incoherence \cite{Rebyakova:2011vf}.  

The incoherent components for the 1n1n and XnXn distributions are fit with a dipole form factor:
\begin{equation}
\frac{d\sigma}{dt} = \frac{A/Q_{0}^{2}}{(1 + |t|/Q_{0}^{2})^{2}}
\end{equation}
which has been used to describe low $Q^{2}$ photon-nucleon interactions \cite{Drees:1988pp}.
The fit is done in the range from $-t = 0.2$ (GeV/c)$^2$ (above the coherent production region) to $-t= 0.45 $ GeV/c$^2$.  The upper limit for $-t$ is chosen to reduce the contamination from hadronic interactions.  For the events with mutual dissociation into any number of neutrons (XnXn), the fit finds
 $A=3.46 \pm 0.02$ mb and $Q_{0}^{2}=0.099 \pm 0.015 \ ({\rm  GeV/c})^{2}$, with  $\chi^{2}/NDF=19/9$. For events with mutual dissociation into single neutrons (1n1n),
 $Q_0^2$ is fixed at 0.099 (GeV/c)$^{2}$.  The fit finds  $A=0.191 \pm 0.003$ mb, with $\chi^{2}/NDF=15.8/10$.  The integrals of these fits lead to the incoherent cross-sections shown in Tab. \ref{tab:ratios}.   The coherent component of the $t$ distribution is then extracted by subtracting the incoherent-component  fit  from the total $d\sigma/dt$.  

\begin{table*}
\begin{tabular} {l|r|r}
Parameter&  XnXn  &  1n1n \\
\hline
$\sigma_{\rm coh.}$ &      $6.49 \pm 0.01 ({\rm stat.}) \pm 1.18 ({\rm syst.})$ mb   &  $0.770 \pm 0.004 ({\rm stat.}) \pm 0.140 ({\rm syst.})$ mb  \\
$\sigma_{\rm incoh.}$&      $2.89 \pm 0.02 ({\rm stat.}) \pm 0.54  ({\rm syst.})$ mb  & $0.162 \pm 0.010 ({\rm stat.}) \pm 0.029  ({\rm syst.})$ mb  \\
\hline
$\sigma_{\rm incoh.}\!/ \!\sigma_{\rm coh.}$& $0.445 \pm 0.015 ({\rm stat.}) \pm 0.005 ({\rm syst.})$  & $0.233 \pm 0.007 ({\rm stat.}) \pm 0.007 ({\rm syst.})$ \\
\end{tabular}
\caption{The coherent and incoherent cross-sections for $\rho^0$ photoproduction with XnXn and 1n1n mutual excitation, and their ratios.} 
\label{tab:ratios}
\end{table*}

If the nuclear excitation was completely independent of $\rho$ photoproduction, then the cross-section ratio for incoherent to coherent production should not depend on the type of nuclear excitation studied.  It is not; the difference could signal the breakdown of factorization, for a couple of reasons.   One possibility is that
unitarity corrections play a role by changing the impact parameter distributions for 1n1n and XnXn interactions.  When $b \gtrsim 2R_A$, the cost of introducing another low-energy photon into the reaction is small.   So, one photon can excite a nucleus to a GDR, while a second photon can further excite the nucleus, leading to Xn emission rather than 1n \cite{Baltz:1996as}.  
The additional photon alters the impact parameter distributions for the 1n1n and XnXn channels.  The XnXn channel will experience a slightly larger reduction at small $|t|$ due to interference from the two production sites.  This may slightly alter the measured slopes and coherent/incoherent ratios.
Alternately, at large $|t|$, a single photon can both produce a \myro and leave the target nucleus excited, breaking the assumed factorization paradigm. The rate has not been calculated for $\rho^0$, but the cross section for $J/\psi$ photoproduction accompanied by neutron emission is significant \cite{Strikman:2005ze}. This calculated $J/\psi$ cross section is noticeably less for single neutron emission than for multi-neutron emission, so $\rho^0$ photoproduction accompanied by neutron emission might alter the XnXn incoherent:coherent cross section ratio more than that of 1n1n.  The difference between the ratios for 1n1n and XnXn collisions is somewhat larger than was found in a previous STAR analysis \cite{Abelev:2007nb}.

\begin{figure}
\includegraphics[width=0.9\columnwidth]{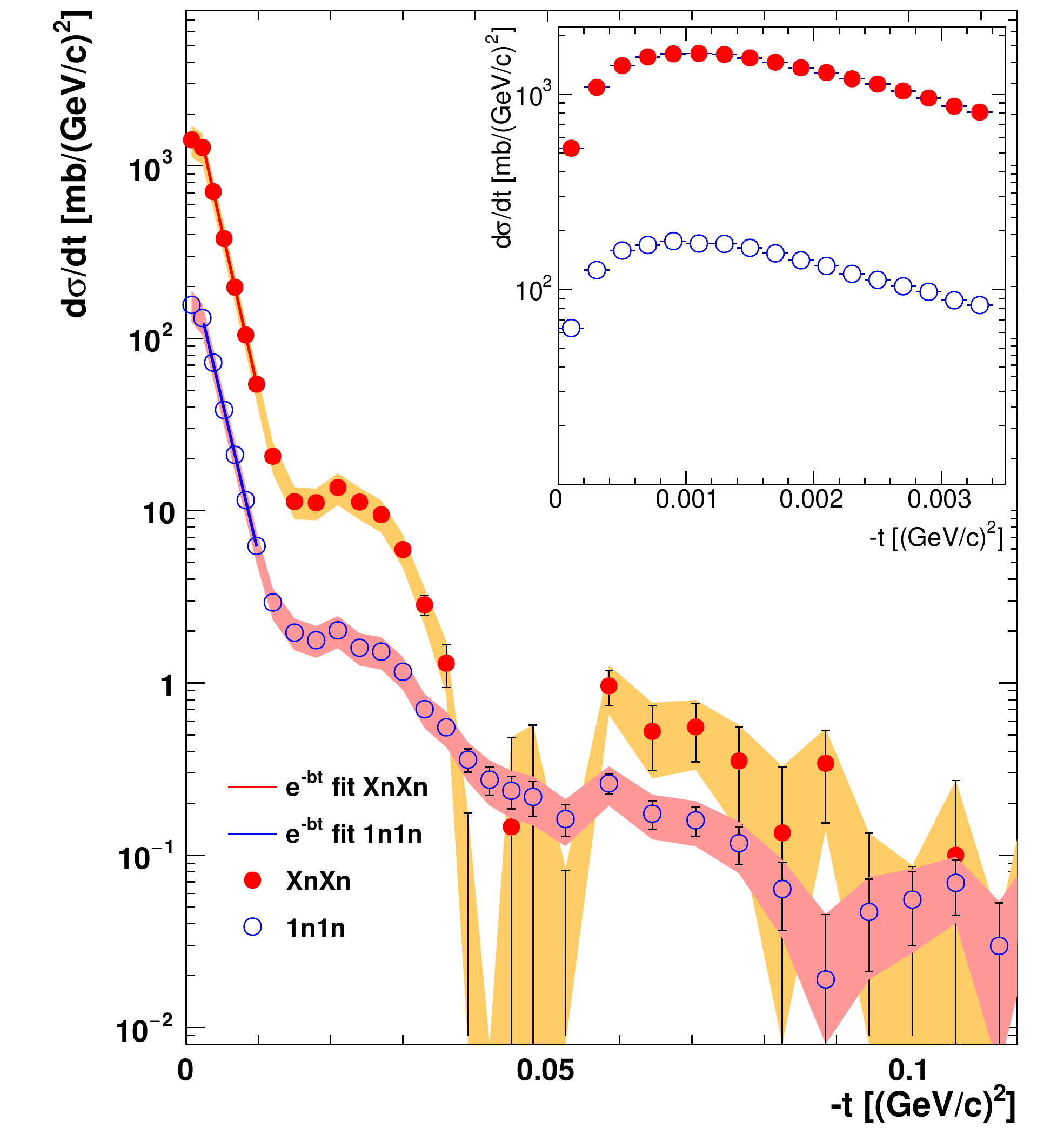}
\caption{\label{fig:dataDiffraction}$d\sigma/dt$ for coherent $\rho^0$ photoproduction in  XnXn events (filled red circles) and 1n1n events (open blue circles).   The filled bands show the sum in quadrature of  all systematic uncertainties listed in Tab. \ref{tab:tableRap} and the statistical errors, which are shown as vertical lines.  The red and blue lines show an exponential fit at low $t$, as discussed in the text.  The insert shows, with finer
binning at low $p_T$, the effects of the destructive interference between photoproduction with the photon emitted by any of the two ions.
}
\end{figure}

The $d\sigma/dt$ for coherent \myro photoproduction accompanied with mutual dissociation of the nuclei into any number of neutrons (XnXn) and only one neutron (1n1n) is shown in Fig. \ref{fig:dataDiffraction} with red and blue markers, respectively. In  both 1n1n and XnXn events, two well-defined minima can  clearly be seen.   In both spectra, the first minima are at $-t = 0.018\pm0.005~ ({\rm GeV}/c)^{2}$. Second minima are visible at $0.043\pm0.01~ ({\rm GeV/c})^{2}$.  To first order, the gold nuclei appear to be acting like black disks, with similar behavior for 1n1n and XnXn interactions.

A similar first minimum may be visible in ALICE data for lead-lead collisions.  Figure 3 of Ref. \cite{Adam:2015gsa} shows an apparent dip in $dN/dp_T$ for $\rho^0$ photoproduction, around $p_T=0.12$ GeV/c ($-t=0.014$ (GeV/c)$^2$).   Lead nuclei are slightly larger than gold nuclei, so the dip should be at smaller $|t|$. 

These minima are shallower than would be expected for $\gamma-A$ scattering, because the photon $p_T$ partly fills in the dips in the $\gamma-A$ $p_T$ spectrum.   There are several theoretical predictions for the locations and depths of these dips. A classical Glauber calculation found the correct depths, but slightly different locations \cite{Klein:1999gv}.  A quantum Glauber calculation  did a better job of predicting the locations of the first minimum \cite{Frankfurt:2015cwa}, although that calculation did not include the photon $p_T$, so missed the depth of the minimum.   However, quantum Glauber calculations which included nuclear shadowing predict that, because of the emphasis on peripheral interactions, the nuclei should be larger, so the diffractive minima are shifted to lower $|t|$ \cite{Guzey:2016qwo}.   For $\rho$ photoproduction with lead at LHC energies, this calculation predicted that the first minima should be at about 0.0165 (GeV/c)$^2$ without the shadowing correction, and 0.012 (GeV/c)$^2$ with the correction.  These values are almost independent of collision energy, but depend on the nuclear radii. Scaling by the ratio of the squares of the nuclear radii, 1.078, the predictions are about 0.0177 (GeV/c)$^2$ without the shadowing correction, and 0.0130 (GeV/c)$^2$ with the shadowing.  The data is in better agreement with the prediction that does not include the shadowing correction.

The Sar\it{t}\rm{r}e event generator run in UPC mode at RHIC energies \cite{Toll:2013gda} produces a Au nucleus recoil after \myro elastic scattering with a very good agreement with the \myro $t$ distribution presented here.  That is not surprising, since it includes a physics model that is similar to the quantum Glauber calculation that does not include nuclear shadowing.

An exponential function is used to characterize the spectrum below the first peak ($0.0024<|t|<0.0098$~ ({\rm GeV}/c)$^2$). The measured slope is $426.4 \pm 1.8~ ({\rm GeV/c})^{-2}$ for the XnXn events and $407.8 \pm 3.2~ ({\rm GeV/c})^{-2}$ for the 1n1n events.     The XnXn slope is very similar to the ALICE measurement of $426\pm 6 \pm 15$ ~({\rm GeV/c})$^{-2}$ \cite{Adam:2015gsa}; there is no evidence for an increase in effective nuclear size with increasing photon energy. 

At very small $-t$, $|t|<10^{-3}$ (GeV/c)$^2$, both cross sections flatten out and turn downward, as can be seen in the insert in Fig. \ref{fig:dataDiffraction}.  This is expected due to destructive interference between $\rho^0$ production on the two nuclear targets \cite{Klein:1999gv,Abelev:2008ew}.

These results are subject to the common uncertainties from Tab. \ref{tab:common}, in addition to the point-to-point uncertainties described above and listed in Tab. \ref{tab:table2}. The yellow and pink bands in Fig. \ref{fig:dataDiffraction} are the sum in quadrature of all systematic uncertainties and statistical errors.

\begin{figure}[t]
\includegraphics[width=0.9\columnwidth]{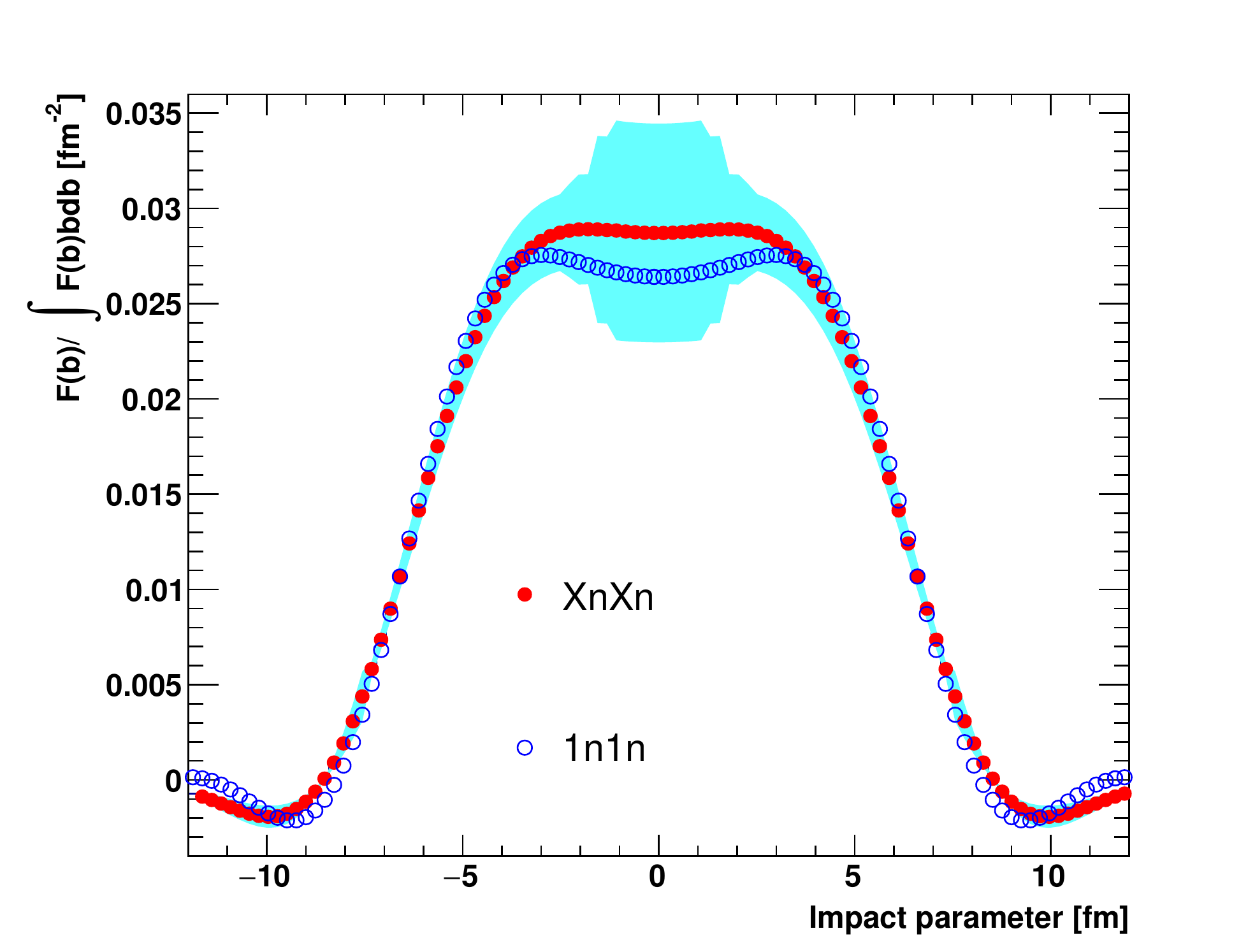}
\caption{\label{fig:fourierTransformation}The target distribution in the transverse plane, the result of a  two-dimensional Fourier transform (Hankel transform) of the XnXn and 1n1n diffraction patterns shown in Fig.
 \ref{fig:dataDiffraction}. The integration is limited to the region $|t| < 0.06$ (GeV/c)$^2$. The uncertainty is estimated by changing the maximum $-t$ to $0.05$, $0.07$ and $0.09$ (GeV/c)$^2$.    The cyan band shows the region encompassed by these $-t$ values.
In order to highlight the similarity of both results at their falling edges, the resulting histograms are scaled by their integrals from -12 to 12 fm.  The FWHM of both transforms is $ 2\times(6.17\pm 0.12)$ fm, consistent with the coherent diffraction of \myro mesons off an object as big as the Au nuclei.}
\end{figure}

The shape of $d\sigma/dt$ for coherent photoproduction is determined by the position of the interaction sites within the target. One can, in principle, determine the density distribution of the gold nucleus via a two-dimensional Fourier transform of $d\sigma/dt$.  RHIC beam energies are high enough that, for $\rho^0$ photoproduction at mid-rapidity, the longitudinal density distribution may be neglected and the ions may be treated as discs.  Nuclei are azimuthally symmetric, so the radial distribution can be determined with a Fourier-Bessel (Hankel) transformation:  
\begin{equation}
F(b) \propto  \frac{1}{2\pi}\int_{0}^\infty dp_T p_T J_0 (bp_T) \sqrt{\frac{d\sigma}{dt}}
\label{Fourier-Bessel}
\end{equation}
Figure \ref{fig:fourierTransformation} shows the result of this transform in the region $|t| <  0.06$ (GeV/c)$^2$. Several features are visible.  The tails of $F(b)$ are negative around $|b|$=\ 10 fm.  This may be due to interference between the two nuclei, since the drop in $d\sigma/dt$ for $|t|<0.0002$ (GeV/c)$^2$ is due to what is effectively a negative amplitude for photoproduction on the `other' nucleus \cite{Abelev:2008ew}. 

We varied the maximum $|t|$ used for the transform over the range $0.05$ to $0.09$ (GeV/c)$^2$.  This led to substantial variation at small $b$, shown by the cyan region in Fig. \ref{fig:fourierTransformation}.   The origin of this variation is not completely clear, but it may be related to aliasing due to the lack of a windowing function \cite{windowing}, or because  of the limited statistics at large $|t|$.  There is much less variation at the edges of the distribution, showing that the transform is stable in the region $4 < b < 7$ fm. The full-width half-maximum (FWHM) of the distribution is $2\times(6.17\pm 0.12)$ fm.
This FWHM is a measure of the hadronic size of the gold nucleus.  With theoretical input, it could be compared with the electromagnetic (proton) radius of gold, as determined by electromagnetic scattering.  The difference would be a measure of the neutron skin thickness of gold, something that is the subject of considerable experimental interest \cite{Tarbert:2013jze,Gardestig:2015eca}.

There are a few effects that need to be considered in comparing the distribution in Fig. \ref{fig:fourierTransformation} with nuclear data.  Because of the significant $q\overline q$ dipole size, $\rho^0$ production occurs preferentially on the front side of the nucleus, and the contribution of the central region is reduced.  Since the photons come from the fields of the other nucleus, the photon field is not uniform across the target; it is stronger on the 'near' side.  Finally, the interference between production on the two targets alters the distributions at large $|b|$.

\section{Summary and Conclusions}

STAR has made a high-statistics study of $\rho^0$, $\omega$ and direct $\pi^+\pi^-$ photoproduction in 200 GeV/nucleon-pair gold-on-gold ultra-peripheral collisions, using 394,000 $\pi^+\pi^-$ pairs.  

We fit the invariant mass spectrum to a mixture of $\rho^0$, $\omega$ direct $\pi^+\pi^-$ and interference terms.  The ratio of direct $\pi^+\pi^-$  to $\rho^0$ is similar to that in previous measurements, while the newly measured $\omega$ contribution is comparable with predictions based on the previously measured $\gamma p\rightarrow \omega p$ cross section and the $\omega\rightarrow\pi^+\pi^-$ branching ratio.  The relative fractions of $\rho^0$, $\omega$, and direct $\pi^+\pi^-$ do not vary significantly with rapidity, indicating that they all have a similar dependence on photon energy.  

We also measure the cross section $d\sigma/dt$ over a wide range, and separate out coherent and incoherent components.  The coherent contribution exhibits multiple diffractive minima, indicating that the nucleus is beginning to act like a black disk.  

This measurement provides a nice lead-in to future studies of photo- and electro-production at an electron-ion collider (EIC) \cite{Accardi:2012qut}, where nuclei may be probed with photons at a wide range of $Q^2$ \cite{Toll:2012mb}.

\section{Acknowledgments}

We thank the RHIC Operations Group and RCF at BNL, the NERSC Center at LBNL, and the Open Science Grid consortium for providing resources and support. This work was supported in part by the Office of Nuclear Physics within the U.S. DOE Office of Science, the U.S. National Science Foundation, the Ministry of Education and Science of the Russian Federation, National Natural Science Foundation of China, Chinese Academy of Science, the Ministry of Science and Technology of China and the Chinese Ministry of Education, the National Research Foundation of Korea, GA and MSMT of the Czech Republic, Department of Atomic Energy and Department of Science and Technology of the Government of India; the National Science Centre of Poland, National Research Foundation, the Ministry of Science, Education and Sports of the Republic of Croatia, RosAtom of Russia and German Bundesministerium fur Bildung, Wissenschaft, Forschung and Technologie (BMBF) and the Helmholtz Association.


\end{document}